\documentclass{svjour3}                     
\smartqed  
\usepackage{graphicx,amsmath}
\newcommand{\blf}[1]{\bf  {\tilde #1}}
\newcommand{\bq}{\begin{eqnarray}}
\newcommand{\eq}{\end{eqnarray}}
\newcommand{\bm}[1] {\mbox{\boldmath{$#1$}}}
\newcommand{\bi}{\begin{itemize}}
\newcommand{\ei}{\end{itemize}}
\def\sumint{\int \! \!\ \! \! \! \! \!\ \! \! \!\! \!\sum}
\def\Tr{\rm Tr}
\newcommand {\vece}[1]{\overset{\rightarrow}{#1}}

\begin{document}

\title{Neutron Transverse-Momentum Distributions and Polarized $^3$He within Light-Front 
Hamiltonian Dynamics
\thanks{Presented at the 20th International IUPAP Conference on Few-Body Problems in Physics, 
20 - 25 August, 2012, Fukuoka, Japan}
}


\author{Emanuele Pace   \and  Giovanni Salm\`e \and  Sergio Scopetta \and Alessio Del Dotto 
\and
Matteo Rinaldi }
\institute{ E. Pace \at
              Universit\`a di Roma ``Tor Vergata'' and INFN, Roma 2, Italy \\
              Tel.: +39-06-72594565\\
              Fax: +39-06-2023507\\
              \email{pace@roma2.infn.it}           
           \and
          Giovanni Salm\`e \at
             INFN Sezione di Roma, Italy
	       \and
          Sergio Scopetta  \at
             Universit\`a di Perugia and
 INFN, Sezione di Perugia, Italy
	       \and
          Alessio Del Dotto \at
              Universit\`a di Roma Tre and INFN, Roma 3, Italy
	       \and
          Matteo Rinaldi \at
              Universit\`a di Perugia and
 INFN, Sezione di Perugia, Italy
}
\date{}
\maketitle
\begin{abstract}
The possibility to extract the quark transverse-momentum distributions in the neutron from
semi-inclusive deep inelastic electron scattering off polarized $^3$He is illustrated 
through an impulse
approximation analysis in the Bjorken limit. The generalization of the analysis
 at finite  momentum transfers in a Poincar\'e
covariant framework is outlined.
The definition of the light-front spin-dependent spectral function of a J=1/2
system allows us to show that within the
light-front dynamics only three of the six leading twist T-even transverse-momentum distributions 
are independent.
\end{abstract}
\vskip -0.2cm
\section{Introduction}
\label{intro}
As is well known, most of the proton spin is carried
 by the quark orbital angular momentum, $L_q$, and by the gluons.
Information on the quark transverse momentum distributions (TMDs) \cite{Barone} and then on $L_q$
can be  accessed through non forward processes, as 
semi-inclusive deep inelastic electron scattering (SIDIS). 
In Ref. \cite{Cates}  
 the possibility
to extract information on the neutron TMDs from 
experimental measurements of the single spin asymmetries (SSAs) on $^3$He was proposed.
In  particular {{SSAs}} allow one to experimentally distinguish the Sivers and the Collins 
asymmetries, 
 expressed in terms of different 
TMDs
and {{fragmentation functions}} (ff) \cite{Barone,Sco}. A large Sivers asymmetry 
was measured in ${{{\vece p}(e,e'\pi)x}}$
\cite{Hermes} and a small one in ${{{\vece D}(e,e'\pi)x}}$ \cite{COMPASS}.
This puzzle has attracted a great interest in obtaining new information on the neutron TMDs.
\section{Neutron properties and a polarized $^3$He target}
A polarized $^3$He is an ideal target 
to study the {{neutron}}, since at a 90\% level a polarized $^3$He is equivalent to
a polarized neutron.
Dynamical nuclear effects in inclusive deep inelastic electron scattering  
 {{$^3\vece{He}( e,e')X$}} (DIS) were evaluated with a realistic 
spin-dependent spectral function 
for $^3{{\vece{He}}}$, ${{ P_{\sigma,\sigma{\prime}} (\vec p, E)}}$ \cite{Ciofi}. It was found  
  that the formula
  \vskip -0.4cm
\begin{equation}
  {{A_n }}\simeq {1 \over 
{{p_n}} f_n} \left 
( {A^{exp}_3} - 2 
{p_p} f_p
{{A^{exp}_p}} \right )~, \quad   \nonumber \\
{\quad \quad \quad \quad \quad \quad 
(f_p, f_n \quad {{dilution factors}})} 
\vspace{-0.1cm}
\end{equation}
can be safely adopted to extract the neutron information from $^3$He and proton data
and it is actually
 used by experimental collaborations.
The nuclear effects are hidden
in the proton and  neutron {{"effective polarizations"}} 
$ {{p_p}} = -0.023  $, 
${{p_n}}= 0.878 $ \cite{Sco}.

To investigate if an analogous formula can be used to extract the {{SSAs}},
  in \cite{Sco}
the process {{$^3\vece{He}( e,e'\pi)X$}} 
was evaluated 
 in the Bjorken limit and
in impulse approximation (IA), i.e.
 the  final state interaction (FSI) was considered only between
 the  two-nucleon 
 which recoil.
 In IA, {{SSAs}} for $^3He$
involve convolutions of 
${{ P_{\sigma,\sigma{\prime}} (\vec p, E)}}$,
 with TMDs
 and  ff. 
Ingredients of the calculations were: i)
a realistic 
${{ P_{\sigma,\sigma{\prime}} (\vec p, E)}}$
for {{$^3$He}} \cite{Kiev},
obtained
using the {{AV18}} interaction
ii) parametrizations of data for TMDs and 
ff, whenever available;
iii) models for the unknown TMDs and 
ff.
As shown in Fig. 1,
in the Bjorken limit the extraction procedure through the  formula
successful in DIS  works nicely 
for the Sivers SSA and the same was shown to occur for the Collins SSA \cite{Sco}.
\begin{figure}
\includegraphics[width=6.3cm]{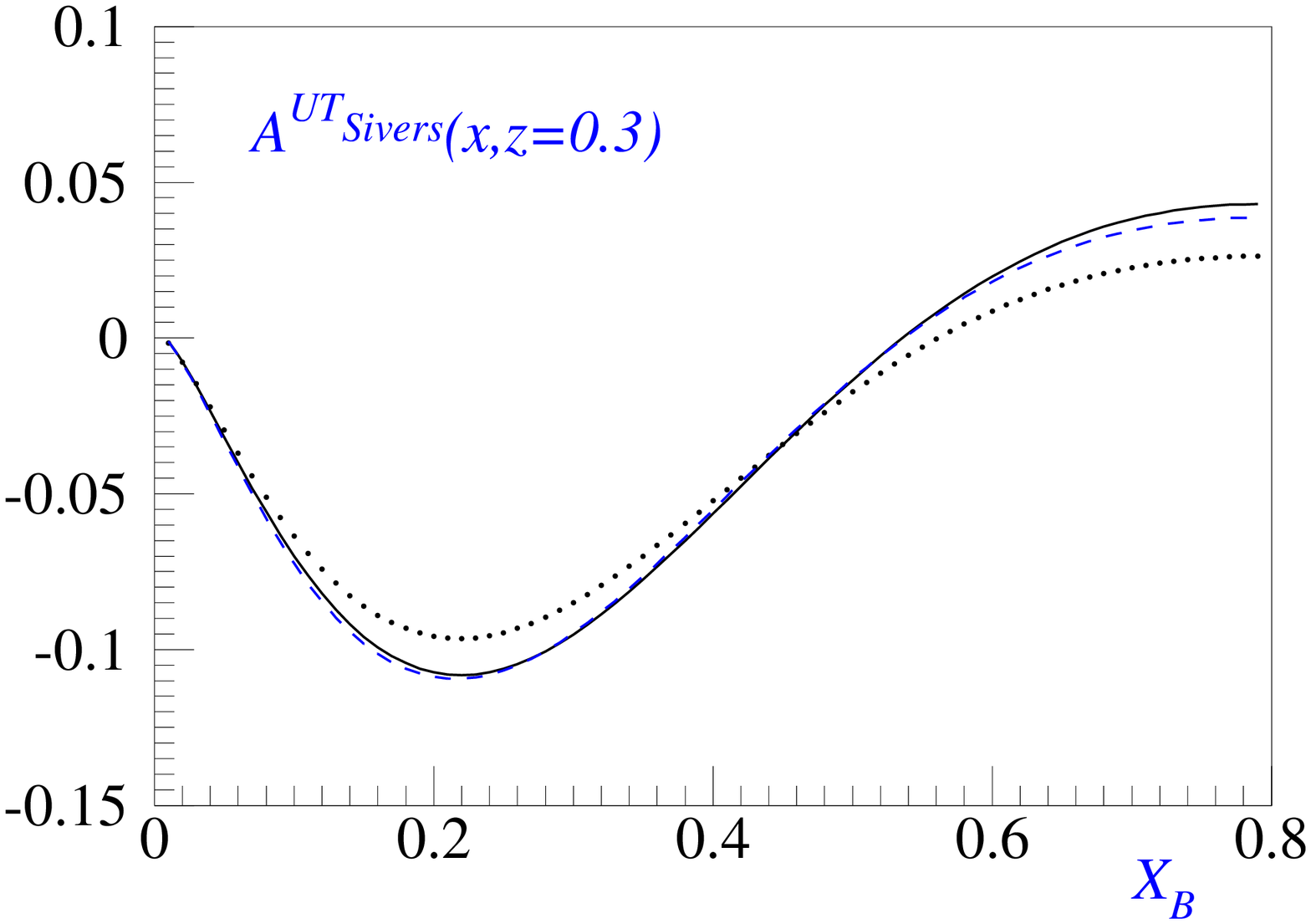}
\includegraphics[width=6.3cm]{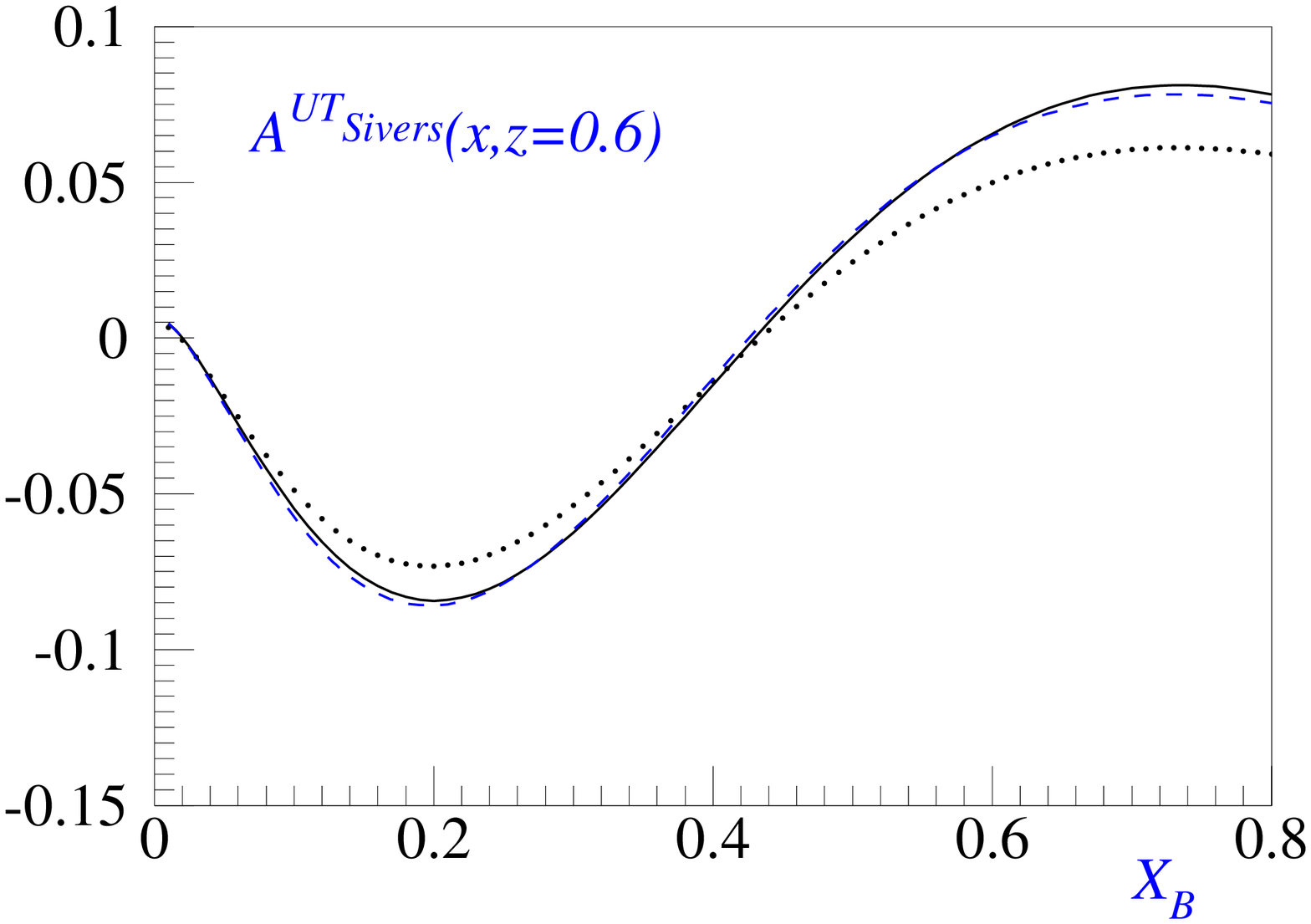}
\vspace{-4cm}
\caption{Sivers asymmetry. Full line: neutron asymmetry of the adopted model;
dotted line: neutron asymmetry extracted 
from the calculated $^3He$ asymmetry neglecting the proton polarization contribution:
 $ {{\bar{A}_{n}}} \simeq {1 \over 
{f_n}} {{A^{calc}_3}}$;
dashed line:
neutron asymmetry extracted 
from the calculated $^3He$ asymmetry taking 
into account  
nuclear structure  through Eq. (1) (after \cite{Sco}).}
\label{fig:1}       
\end{figure}

In \cite{Sco} the calculation was performed in the Bjorken limit.
To study relativistic effects in the 
actual experimental kinematics we adopted \cite{Dotto} the light-front (LF) form of 
Relativistic Hamiltonian Dynamics (RHD) introduced by 
Dirac.
Indeed the {{RHD} of an interacting system with a fixed number of on-mass-shell constituents, 
{\it plus} the Bakamijan-Thomas 
construction of the Poincar\'e generators 
allow one to generate a description of SIDIS off $^3$He
which 
is fully Poincar\'e covariant.

 

In IA the {LF hadronic tensor} for the $^3$He nucleus is:
 \vskip -0.4cm
\begin{eqnarray}
{{ 
{\cal W}^{\mu\nu}(Q^2,x_B,z,\tau, \hat{\bf h},S_{He})}} 
 \propto
 \sum_{\sigma,\sigma'}\sum_{\tau} 
 \left.\sumint \right._{\epsilon^{min}_S}^{\epsilon^{max}_S}{~d
\epsilon_{S} }
\int_{M^2_N}^{(M_X-M_S)^2} dM^2_f \quad \quad
\\
 \times 
\int_{\xi_{l}}^{\xi_{u}} {d\xi\over \xi^2 (1-\xi)(2\pi)^3}
\int_{P^{m}_\perp}^{P^{M}_\perp}{d
P_\perp\over sin\theta   }~ (P^++q^+- h^+)
~ 
{{
w^{\mu\nu}_{\sigma\sigma'}\left(\tau,{\blf q},{\blf h},{\blf P}\right)
}}
{{
{\cal P}^{\tau}_{\sigma'\sigma}(\tilde{\bf k},\epsilon_S,S_{He})
}}
\nonumber
\end{eqnarray}
where $ \tilde{\bf v} = \{v^+=v^0+v^3, {\bf v_{\perp} } \}$,
{$ w^{\mu\nu}_{\sigma\sigma'}
\left(\tau,{\blf q},{\blf h},{\blf P}\right)$
is the nucleon hadronic tensor} and
{{
$
{\cal P}^{\tau}_{\sigma'\sigma}(\tilde{\bf k},\epsilon_S,S_{He})
$
 the {{LF}} nuclear spectral function 
given in terms of the unitary Melosh Rotations, 
$
{{D^{{1 \over 2}} [{\cal R}_M ({\blf k})]}}$,
}}
and the {{instant-form spectral function}}
$
{{
{\cal S}^{\tau}_{\sigma'_1\sigma_1}({\bf k},\epsilon_{S},S_{He})
}}$:
\bq
{ {
  {\cal P}^{\tau}_{\sigma'\sigma}({\blf k},\epsilon_{S},S_{He})
}}
\propto
~\sum_{\sigma_1 \sigma'_1} 
{{D^{{1 \over 2}} [{\cal R}_M^\dagger ({\blf
k})]_{\sigma'\sigma'_1}}}~
{{
{\cal S}^{\tau}_{\sigma'_1\sigma_1}({\bf k},\epsilon_{S},S_{He})
}} ~
{{D^{{1 \over 2}} [{\cal R}_M ({\blf k})]_{\sigma_1\sigma}}}
\eq
 \vskip -0.2cm
Notice that
$
{\cal S}^{\tau}_{\sigma'_1\sigma_1}
$
is given in terms of {{three independent functions}},
${{B_{0},B_{1},B_{2}}}$ \cite{Kiev}.

We are now  evaluating the SSAs using the {{LF hadronic tensor}},
at finite values of $Q^2$. 
%
The preliminary results are quite encouraging, since, as shown in Table 1,
{{LF}} longitudinal and transverse polarizations weakly differ and  
we find that
in the Bjorken limit the extraction procedure works well within 
{{the LF approach}} as it does in the non
relativistic case.
Furthermore the effect of the finite {{integration limits}}
in the actual JLAB kinematics \cite{Qian}, instead of the ones in the Bjorken limit,
is small and
 will be even smaller
in the JLAB planned experiments at 12 GeV \cite{Cates}.

Concerning the FSI, we plan to include the FSI
between the jet produced from the hadronizing quark
 and the two nucleon system through a Glauber approach \cite{Kaptari}.
\vskip 0.2cm

\begin{tabular}{|c|c|c|c|c|c|}
\hline  &\hspace{-1mm}$proton \, {NR}$\hspace{-2mm}&\hspace{-1mm}$proton \,
{LF}$\hspace{-2mm}&\hspace{-1mm}$neutron \, {NR}$\hspace{-2mm}&\hspace{-1mm}$neutron \, {LF}$\hspace{-1mm} \\ 
\hline 
\hspace{-2mm} $\int dE d\vece{p}\,\frac{1}{2}Tr( {\cal{P}} \sigma_{z})_{\vec{S}_A=
\widehat{z}}$ \hspace{-2mm} & -0.02263 & {{-0.02231}} & 0.87805 & 
{{0.87248}} \\ 
\hline 
\hspace{-2mm} $\int dE d\vece{p}\,\frac{1}{2}Tr( {\cal{P}} \sigma_{y})_{\vec{S}_A=
\widehat{y}}$ \hspace{-2mm}
& -0.02263 & {{-0.02268}} & 0.87805 & {{0.87494}} \\ 
\hline 
\end{tabular} 
\section{The $J =1/2$ {{LF}} spectral function and the nucleon {{LF}} TMDs}
The TMDs for a $J =1/2$ system are introduced
through the {{q-q correlator}}
\begin{eqnarray}
  {{\Phi(k, P, S)}}_{\alpha\beta} =
   \int {d^4z} ~ e^{i k{\cdot}z}
 \langle P S | ~\bar \psi_{q\beta}
(0) ~  \psi_{q\alpha} (z) | P S
\rangle \nonumber 
   = \frac12
  \left\{ \phantom{\frac1M} \hspace{-4mm} 
    {{A_1}} 
\, {P}\hspace{-2mm} / \hspace{1mm} +
    { {A_{2}}} 
\, S_L \, \gamma_5 \, {P}\hspace{-2mm} / \hspace{1mm} +
  \right.
  \nonumber
\\
   \null 
 \hspace{-2mm} \left.
  { {A_3}} 
\, {P}\hspace{-2mm} / \, \gamma_5 \, {S}_\perp\hspace{-4mm} / ~~+   \frac1{M} \, 
{{\widetilde{A}_1}} 
\, \vece{k}_\perp{\cdot}\vece{S}_\perp \,
    \gamma_5 {P}\hspace{-2mm} /
  \hspace{1mm}  + \,
{{\widetilde{A}_2}} 
\, \frac{S_L}{M} \,
    {P}\hspace{-2mm} / \, \gamma_5 \, {k}_\perp\hspace{-4mm} /
  \right.
  \left.
\hspace{2mm} +  \, \frac1{M^2} \, 
{{\widetilde{A}_3}} 
\, \vece{k}_\perp{\cdot}\vece{S}_\perp \,
    {P}\hspace{-2mm} / \, \gamma_5 \, {k}_\perp\hspace{-4mm} / \hspace{2mm}
  \right\}_{\alpha\beta} ,~~
\end{eqnarray}
so that the {{six twist-2 T-even TMDs}}, {{$A_i,~
\widetilde{A}_i ~ (i=1,3)$}}, can be obtained by proper traces of 
{{$\Phi(k, P, S)$}}.
 Let us consider the  contribution to the {{correlation function}}
from  on-mass-shell fermions
\vskip -0.6cm
\bq
{ {~\Phi_p(k,P,S)=~{(~{  k \hspace{-2mm} /}_{on}~ + ~m )\over 2 m}~
{{\Phi(k,P,S)}}~{(~{  k\hspace{-2mm} /}_{on}~ + ~m )\over 2 m} }}=
\\   = 
\sum_\sigma \sum_{\sigma'}~u_{LF}({\tilde k},\sigma')~\bar{u}_{LF}( {\tilde k},\sigma')
~{{\Phi(k,P,S)}}
~u_{LF}({\tilde k},\sigma)\bar{u}_{LF}( {\tilde k},\sigma) \nonumber
\hspace{-0.3mm}
\eq 
and let us identify $\bar{u}_{LF}( {\tilde k},\sigma')
~\Phi(k,P,S)
~u_{LF}({\tilde k},\sigma)$ 
with 
the {{LF nucleon spectral function}},
${\cal P}^{}_{\sigma'\sigma}(\tilde{\bf k},\epsilon_S,S)$.
In a reference frame where ${\bf P}_\perp = 0$, the following relation holds
between the off-mass-shell minus component {$k^-$} of the struck quark and 
the spectator diquark energy {{$\epsilon_S $}} :
\vskip -0.3cm
\bq
{{k^-}} = {{M^2}\over {P^+}} ~ - ~ {{({{\epsilon_S}} + m) ~ 4m + |{\bm k}_\perp|^2}\over {P^+ -k^+}}
\eq
The  
 traces of $~[\gamma^+ ~\Phi_p(k,P,S)]$,~ $[\gamma^+ ~\gamma_5~~\Phi_p(k,P,S)]$, and
 $[{k}\hspace{-2mm} / _\perp\gamma^+
\gamma_5~\Phi_p(k,P,S)]$ can be obtained in terms of the TMD's, {{$A_i,~
\widetilde{A}_i ~ (i=1,3)$}}, through Eq. (4). 
However these same traces can be also expressed through the {{LF spectral function}},
since
\bq
{ {{1 \over 2 P^+}~ Tr\left[ \gamma^+ ~\Phi_p(k,P,S) \right]}} ~ 
=
~{k^+\over 2m P^+} ~  { {Tr\left[ {\cal P}^{}_{}(\tilde{\bf k},\epsilon_S,S)
\right]}}
\eq
\vspace{-6mm}
\bq
{ {{1 \over 2 P^+}~Tr\left[ \gamma^+ ~\gamma_5~~\Phi_p(k,P,S)\right]}}=
 ~{k^+\over {2m P^+}} ~  { {
Tr\left[ \sigma_z ~ {\cal P}^{}(\tilde{\bf k},\epsilon_S,S) \right]}}
\eq
\vspace{-6mm}
\bq
{ {{1 \over 2 P^+}
Tr\left[ {k}\hspace{-2mm} / _\perp\gamma^+
\gamma_5~\Phi_p(k,P,S)\right]}} 
={k^+\over {2m P^+}} ~  {{
Tr\left[ {\bm k}_\perp \cdot {\bm \sigma} ~ 
{\cal P}^{}(\tilde{\bf k},\epsilon_S,S) \right]}} \quad \quad
\eq
In turn the traces 
${1 \over 2} {\Tr}( {\cal P} I)$, 
${1 \over 2} {\Tr}( {\cal P} \sigma_z )$, 
${1 \over 2}
{\Tr}(  {\cal P} \sigma_i)$ ($i=x,y$) can be expressed in terms of three scalar functions, as in the
$^3$He case,
and known
kinematical factors.
Then in the {{LF approach}} with a fixed number of particles the six 
leading twist TMDs, $A_i,
\widetilde{A}_i ~ (i=1,3)$, 
can be expressed in terms of the previous three independent scalar functions. 
\vspace{-2mm}
\section{Conclusion}
A realistic study of {{ $^3\vece{He}(e,e'\pi)X$ }}
in the Bjorken limit was performed 
in {{IA}}. Nuclear effects
in the extraction of the {{neutron}} information were 
found to be {{under control}}.
An analysis at finite $Q^2$  with a {LF} spectral function
is in progress in order to test the extraction procedure of the neutron 
information from $^3\vece{He}(e,e'\pi)X$ experiments.
The relations found among the six leading twist T-even TMDs from general properties 
{{within LF dynamics and a fixed number of degrees of freedom}}
show that {{only three}}
  of the six {{T-even TMDs are independent}}.}
 These relations are precisely predicted 
within {{LF}} dynamics, 
and could be experimentally
checked to test the {{LF}} description of SIDIS in the valence region.



\end{document}